\documentclass[%
 reprint,
 amsmath,amssymb,
 aps,
 twocolumn,superscriptaddress
]{revtex4-2}

\usepackage{graphicx}
\usepackage{dcolumn}
\usepackage{bm}
\usepackage{xcolor} 
\usepackage{array}
\usepackage[colorlinks=true,linkcolor=blue,citecolor=blue,urlcolor=blue]{hyperref}

\newcommand{\IQA}{\affiliation{1}{International Quantum Academy, Shenzhen 518048, China}}
\newcommand{\SUSTECH}{\affiliation{2}{Southern University of Science and Technology, Shenzhen 518055, China}}
\newcommand{\PKU}{\affiliation{3}{School of Physics, Peking University, Beijing 100871, China}}
\newcommand{\HFNL}{\affiliation{4}{Shenzhen Branch, Hefei National Laboratory, Shenzhen 518048, China}}

\begin{document}

\preprint{APS/123-QED}

\title{Real-time Surface-Code Error Correction Using an FPGA-based Neural-Network Decoder}

\author{Xiaohan Yang$^{*}$}
\thanks{These authors contributed equally to this work.}
\affiliation{\IQA}\affiliation{\SUSTECH}

\author{Xuandong Sun$^{*}$}
\email{sunxd@iqasz.cn}
\affiliation{\IQA}\affiliation{\SUSTECH}

\author{Zhiyi Wu$^{*}$}
\affiliation{\IQA}\affiliation{\PKU}

\author{Jiawei Zhang}
\affiliation{\IQA}\affiliation{\SUSTECH}

\author{Ji Jiang}
\affiliation{\IQA}

\author{Xiayu Linpeng}
\affiliation{\IQA}

\author{Yuxuan Zhou}
\affiliation{\IQA}

\author{Ji Chu}
\email{jichu@iqasz.cn}
\affiliation{\IQA}

\author{Jingjing Niu}
\affiliation{\IQA}\affiliation{\HFNL}

\author{Youpeng Zhong}
\affiliation{\IQA}\affiliation{\HFNL}

\author{Song Liu}
\affiliation{\IQA}\affiliation{\HFNL}

\author{Dapeng Yu}
\email{yudapeng@iqasz.cn}
\affiliation{\IQA}\affiliation{\HFNL}

\date{\today}

\begin{abstract}

Quantum error correction (QEC) is essential for achieving low error rates required for fault-tolerant quantum computation. In stabilizer-based codes such as the surface code, errors are inferred from repeated syndrome measurements and corrected by a classical decoder. 
To prevent error accumulation, decoding must be performed with both high throughput and low latency to keep pace with the QEC cycle and enable real-time feedback for universal logical operations.
Here we report a hardware-integrated control architecture featuring an FPGA-based neural-network (NN) decoder and experimentally demonstrate real-time surface-code (distance-3) QEC on a superconducting quantum processor. The system achieves a deterministic closed-loop latency of 550~ns, including 124~ns for NN decoding, enabling feedback corrections within a 1.25~us QEC cycle. We show that real-time decoding and feedback correction achieve logical performance comparable to offline decoding while maintaining robustness against varying error conditions. We further demonstrate mid-circuit feedback correction in non-Clifford logical circuits, where Pauli-frame updating alone becomes insufficient. Our results establish a low-latency hardware architecture for embedded QEC control and provide a pathway towards scalable fault-tolerant quantum computing systems.

\end{abstract}

\maketitle

\section{introduction}

\begin{figure*}[htbp]
  \centering
  \includegraphics[width=\textwidth]{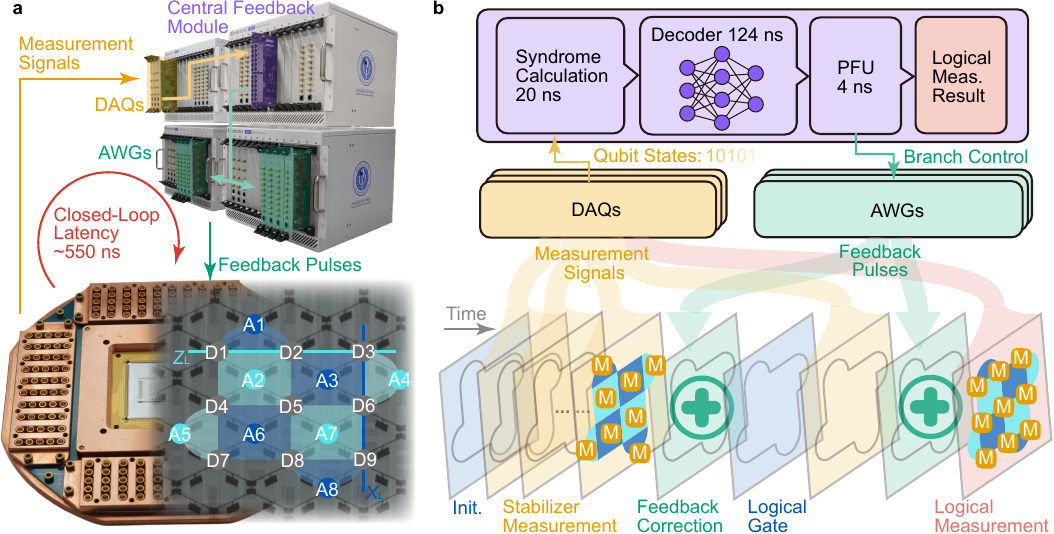}
  \caption{\textbf{System architecture and real-time QEC workflow.}   
    \textbf{a}, Schematic of the experimental setup, showing in-house room-temperature control electronics connected to a 66-qubit superconducting quantum processor operating at cryogenic temperatures. The control electronics integrate data acquisitions (DAQs, orange) for acquiring measurement signals, arbitrary waveform generators (AWGs, green) for generating qubit control pulses, and a central feedback module (CFM, purple) with a NN decoder. A 17-qubit subset is used to implement the distance-3 surface code: data qubits (D1--D9) store the logical information, while ancilla qubits (A1--A8) perform \(Z\) (light blue) and \(X\) (dark blue) stabilizer measurements.
    \textbf{b}, Real-time QEC workflow. During each stabilizer measurement, measurement signals are acquired and processed by the DAQs to extract ancilla states. The resulting bit strings are transmitted to the CFM, where error syndromes are computed (20~ns latency) and decoded by the NN decoder (124~ns). 
    The decoding output is used to update the Pauli frame (4~ns) and trigger conditional feedback pulses via the AWGs.
    The overall closed-loop latency—from the end of the readout pulse to the start of the feedback pulse—is \(550\,\mathrm{ns}\), as indicated in (a).
    In addition to the feedback loop, logical operations (blue) and final logical measurements (red) on the data qubits are incorporated into the sequence to complete the QEC workflow.
    }
  \label{fig:fig1}
\end{figure*}

Quantum computing promises transformative speedups for problems in simulation~\cite{lloyd1996universal}, optimization~\cite{abbas2024challenges}, and cryptography~\cite{shor1999polynomial}. 
Achieving these advantages at scale requires robust mitigation of noise and control imperfections in physical qubits, which makes quantum error correction (QEC) essential~\cite{shor1995scheme}.
Stabilizer codes~\cite{gottesman1998theory}, including surface code~\cite{fowler2012surface} and color codes~\cite{bombin2006topological}, have emerged as leading approaches to fault-tolerant quantum computation, with recent experiments demonstrating reduced logical error rates with increasing code distance~\cite{google2023suppressing,google2025quantum,he2025experimental,lacroix2025scaling,bluvstein2026fault}. 
QEC requires repeated stabilizer measurements to extract error syndromes, which are processed by a classical decoder to infer and correct errors~\cite{shor1995scheme,steane1996Error,gottesman1998theory}. 
The associated classical operations must be performed in real time~\cite{battistel2023realtime,skoric2023parallel}, with two key performance metrics: throughput and closed-loop latency.
Throughput sets the rate at which syndromes are processed: if the decoder cannot keep pace with syndrome generation, a backlog of classical data can slow or stall the QEC cycle, undermining fault tolerance and the potential quantum speedup~\cite{terhal2015Quantuma}. 
Closed-loop latency—including decoding, communication, and feedback delays—determines how quickly corrective operations can be applied, which is particularly important for non-Clifford gates, where Pauli-frame updating (PFU) alone is insufficient~\cite{bravyi2005universal,litinski2019game}.

Achieving real-time error correction for superconducting circuits with fast QEC cycles ($\sim1\mu$s) is extremely challenging, requiring both accurate low-latency decoding and tight integration with the control hardware.
Existing hardware implementations generally fall into two categories.
GPU-, CPU-, and TPU-based systems support computationally intensive decoding algorithms with strong parallelism~\cite{huang2020fault,skoric2023parallel,tan2023Scalable,higgott2025sparse}, particularly neural-network (NN) decoders that achieve high throughput at large code distances~\cite{baireuther2019neural,sweke2021reinforcement,meinerz2022scalable,gicev2023scalable,bausch2024learning,senior2025scalable,varbanov2025neural,lee2025Scalable,zhang2026learning,chamberland2026fast}. However, communication overhead and nondeterministic execution latency hinder deterministic real-time feedback~\cite{battistel2023realtime,caldwell2025platform,muller2025improved}.
By contrast, FPGA- and ASIC-based systems~\cite{valls2021SyndromeBased,overwater2022neuralnetwork,das2022AFS,liyanage2023Scalablea,liao2023WITGreedy,liyanage2024fpgabased,caune2024demonstrating,barber2025realtime,maurya2025fpgatailored,maurer2025realtime,kadomoto202522nm,wu2025Micro,ziad2025Local,bascones2025Exploring,zhang2025LATTE,maan2026decoding} provide deterministic timing and direct integration with quantum-control electronics, although their limited hardware resources constrain decoder complexity.
Despite substantial progress in low-latency decoding, realizing a real-time, feedback-corrected logical qubit using a fully integrated control system remains an outstanding challenge.

In this work, we develop a hardware-integrated control architecture featuring FPGA-based NN decoding and experimentally demonstrate real-time surface-code (distance-3) QEC on a superconducting quantum processor. 
The NN decoder achieves a latency of 124~ns with a throughput period of 184~ns, and the complete system realizes a deterministic closed-loop latency of 550~ns, enabling feedback within a sub-microsecond QEC cycle. 
We show that real-time decoding combined with feedback correction achieves logical performance and robustness comparable to those of an offline minimum-weight perfect matching (MWPM) decoder. Furthermore, we demonstrate real-time error correction in non-Clifford logical circuits, where PFU or post-processing alone is insufficient because Pauli errors propagate beyond the Pauli frame.
These results establish a deterministic, hardware-integrated feedback architecture for low-latency QEC, supporting future implementations of universal fault-tolerant quantum computation.

\begin{figure*}[htbp]
  \centering
  \includegraphics[width=0.7\textwidth]{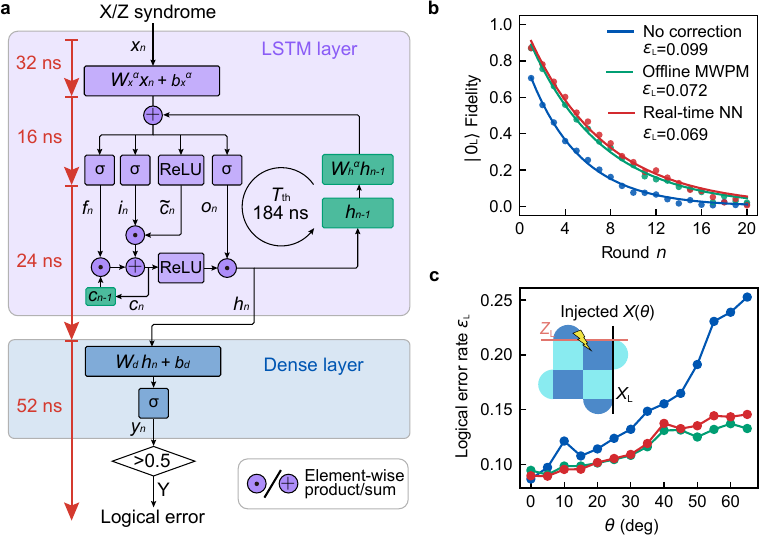}
  \caption{\textbf{FPGA-integrated neural-network decoder.}  
  \textbf{a}, FPGA implementation of the recurrent neural-network decoder. X-type and Z-type decoders process syndrome inputs \(x_n\) from X and Z stabilizers, respectively, and output logical flip instructions. The LSTM layer maintains temporal memory via recurrent cell state \(c_n\) and hidden state \(h_n\), using gating mechanisms (\(i_n, f_n, \tilde{c}_n, o_n\)). Pre-trained weight matrices (\(W_x^\alpha, W_h^\alpha, W_d\), \(\alpha \in \{i,f,c,o\}\)) and bias vectors (\(b_x^\alpha, b_d\)) are quantized to 6-bit integers for efficient FPGA implementation. The critical path comprises four pipelined stages (red arrows), while memory updates (green) run in parallel, preserving low feedback latency. 
  \textbf{b}, Fidelity of \(|0_{\rm L}\rangle\) versus QEC rounds \(n\), comparing real-time NN decoding (red), offline MWPM decoding (green), and uncorrected data (blue). Each point is averaged over 10,000 shots.
  \textbf{c}, Decoder robustness under injected physical errors on D2.}
  \label{fig:fig2}
\end{figure*}

\section{The architecture}

We perform the experiments using in-house developed control electronics~\cite{zhangZhang2024m2} and a 66-qubit superconducting quantum processor~\cite{sun2025logical}, as shown in Fig.~\ref{fig:fig1}a. 
A subset of 17 physical qubits, including 9 data qubits (D1--D9) and 8 ancilla qubits (A1--A8), is used to encode a distance-3 surface-code logical qubit. See the Supplementary Information for device details~\cite{sm2026}.
The control electronics consist of three main components: (1) arbitrary waveform generators (AWGs) for generating control pulses; (2) data acquisition modules (DAQs) for acquiring and processing measurement signals; and (3) a central feedback module (CFM) that implements decoding and feedback logic. 
The CFM contains an AMD Kintex-7 FPGA (XC7K410T) for syndrome decoding, along with 12 communication interfaces for receiving qubit-state data from the DAQs and 12 additional interfaces for sending branch-control signals to backplanes of the AWG chassis. Each backplane also contains an FPGA that distributes the branch-control signals to the AWGs. 

The workflow for real-time QEC is shown in Fig.~\ref{fig:fig1}b.
A typical experiment includes logical state initialization, repetitive stabilizer measurements, feedback correction, logical gates, and a final logical measurement performed on the data qubits.   
During stabilizer measurements or logical measurement, the DAQs acquire and discriminate qubit signals, delivering the resulting bit strings to the CFM via low-latency links. The CFM converts these data into error syndromes in 20~ns and decodes them using the NN decoder with a deterministic latency of 124~ns.  
The decoding outputs are used to update the Pauli frame within the FPGA and to generate branch-control instructions for the AWGs according to the quantum circuit schedule. During logical measurement, the CFM combines the data-qubit readout outcomes with the Pauli frame to produce corrected logical results.  
The total closed-loop latency—from the end of the readout pulses to the start of the feedback pulses—is 550 ns. This includes 222 ns for DAQ sampling after readout pulses, 148 ns for CFM processing, and 180 ns from additional electronic delays (see Methods for a detailed breakdown). 
Consequently, the architecture implements a fully integrated real-time QEC loop in which syndrome calculation, NN decoding, PFU, and conditional feedback correction are all executed entirely within FPGA hardware.

\section{Neural-Network decoder}

The NN decoder comprises an LSTM layer~\cite{hochreiter1997Long,lecun2015Deep} followed by a dense output layer, as shown in Fig.~\ref{fig:fig2}a. At QEC round \(n\), the LSTM processes syndrome data \(x_n\) to produce a hidden state \(h_n\), which is mapped by the dense layer to an output \(y_n\); \(y_n > 0.5\) signals a logical error. 
Temporal correlations across QEC rounds are captured by the LSTM, configured with 32 hidden units ($h_n \in \mathbb{R}^{32}$) to minimize hardware latency without sacrificing decoding accuracy. 
The FPGA implementation employs a four-stage pipelined structure, with parallelized matrix computations on DSP (Digital Signal Processing) blocks to reduce latency. The critical path from input to output has a fixed latency of 124~ns at a 250~MHz clock rate, and the minimum throughput period $T_{\rm th}$ is 184~ns, substantially shorter than a typical QEC cycle.
Separate NN decoders are trained for \(X\)-type and \(Z\)-type stabilizers using STIM-generated~\cite{gidney2021Stim} datasets and fine-tuned with experimental data.
For FPGA deployment, weights are quantized to 6-bit integers, and simplified activation functions are employed for computational efficiency. Both quantization and FPGA implementation introduce negligible performance degradation~\cite{sm2026}. 

\begin{figure*}[htbp]
  \centering
  \includegraphics[width=\textwidth]{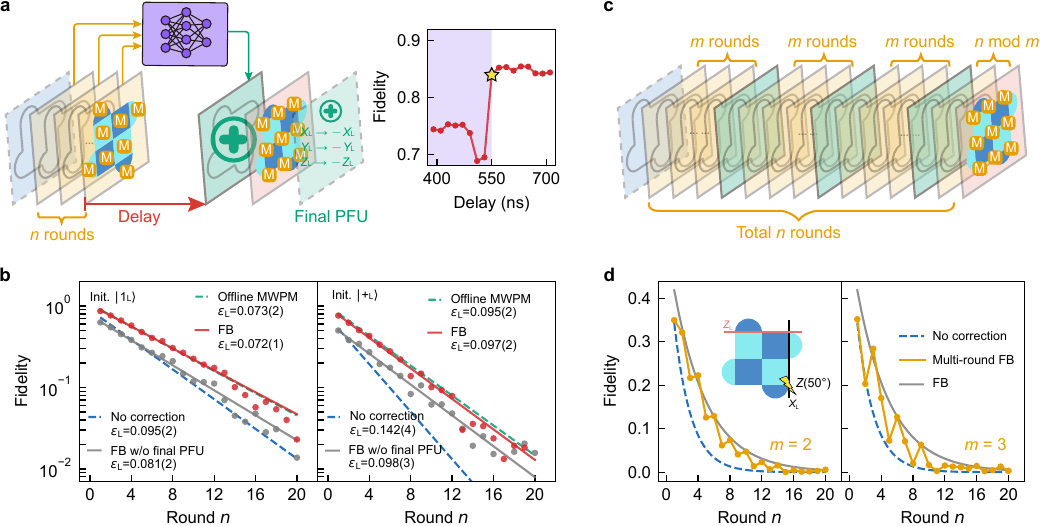}
  \caption{\textbf{Real-time QEC.}
\textbf{a}, Logical memory circuit with repeated stabilizer measurements and real-time feedback. A delay is inserted after each measurement to accommodate the feedback latency, and a final PFU corrects residual errors detected during logical measurement. The right panel shows logical \(|0_{\rm L}\rangle\) fidelity (inset, \(n=1\)) versus the delay time. Delays shorter than the closed-loop latency (550~ns) prevent the decoder output from reaching the AWG in time, corresponding to an infeasible operating regime (purple).
\textbf{b}, Logical-state fidelities as a function of QEC rounds \(n\) with final-round feedback. Results are shown with (red) and without (grey) the final PFU, compared with uncorrected data (blue dashed) and offline MWPM decoding (green dashed).  
\textbf{c}, Logical memory circuit with repeated mid-circuit feedback, where a feedback operation is applied every \(m\) rounds. No final PFU is applied after logical measurement to isolate the effect of the feedback pulses. 
\textbf{d}, Experimental results for \(|+_{\rm L}\rangle\) with periodic feedback (orange), compared with final-round-only feedback (grey) and uncorrected data (blue dashed) as references. Artificial \(Z\)-rotation errors are injected on qubit D9 to highlight the effect of mid-circuit feedback.
}
\label{fig:fig3}
\end{figure*}

We evaluate the accuracy of the NN decoder in a distance-3 surface-code logical memory experiment. Logical errors are corrected directly on the FPGA as sign-bit flips in the PFU registers (e.g., \(+X_{\rm L} \rightarrow -X_{\rm L}\)).
Logical state fidelity \(|0_{\rm L}\rangle\) as a function of QEC rounds \(n\) is shown in Fig.~\ref{fig:fig2}b, compared with offline MWPM decoding~\cite{higgott2021pymatching,higgott2023sparse}.
Fitting fidelity decay to \(F(n) = (1 - 2\epsilon_{\rm L})^n\) yields \(\epsilon_{\rm L} = 6.9(2)\%\) for the NN decoder, comparable to \(\epsilon_{\rm L} = 7.2(2)\%\) for offline MWPM.
To test robustness of the NN decoder under different error conditions, we artificially inject errors by applying \(X(\theta)\) rotations on qubit D2 before each stabilizer measurement (Fig.~\ref{fig:fig2}c). Subsequent \(Z\)-stabilizer measurements stochastically discretize the rotation into logical bit-flip errors, with probability increasing with \(\theta\)~\cite{gutierrez2016Errorsa}. The NN decoder maintains performance comparable to MWPM even at elevated error rates, demonstrating its reliability for real-time QEC.

\section{Real-time QEC}

We demonstrate real-time error correction using the circuit shown in Fig.~\ref{fig:fig3}a. Syndrome data acquired during repeated stabilizer measurements are processed by the NN decoder, updating the Pauli frame. Decoder-triggered feedback corrections are applied before the logical measurement, and a final PFU is performed to correct residual errors detected during logical measurement.
Logical errors are corrected via physical \(X\) or \(Z\) gates on representative data qubits (D1 or D9)~\cite{caune2024demonstrating}, rather than directly applying logical \(X_{\rm L}\) or \(Z_{\rm L}\) gates. These corrections flip neighboring stabilizer outcomes in the subsequent round, and the CFM updates the affected outcomes accordingly.

To determine the minimum closed-loop latency for mid-circuit feedback, we sweep the delay between stabilizer measurement and feedback application while monitoring logical fidelity (right panel of Fig.~\ref{fig:fig3}a). 
Fidelity decreases when the delay is below 550 ns, as the decoder output cannot reach the AWG in time, and recovers for longer delays. We therefore set 550 ns as the operating point. This latency remains well below the QEC cycle time and is sufficiently short to suppress the accumulation of delay-induced logical errors.

Fig.~\ref{fig:fig3}b shows the logical fidelities of \(|1_{\rm L}\rangle\) and \(|+_{\rm L}\rangle\) versus QEC rounds for different correction schemes. With real-time feedback, the logical error rates are 0.072(1) and 0.097(2) per round for \(|1_{\rm L}\rangle\) and \(|+_{\rm L}\rangle\), respectively, closely matching the results obtained via post-processing (0.073(2) and 0.095(2)).
Removing the final PFU yields similar logical error rates, confirming that the improvements primarily arise from the feedback pulses, while the final PFU enhances overall fidelity by correcting residual errors detected in the final measurement.

We further demonstrate repeated multi-round feedback which enables immediate correction after each stabilizer measurement. 
Feedback operations are periodically inserted throughout the circuit (Fig.~\ref{fig:fig3}c), and artificial \(Z\) errors are injected on qubit D9 prior to each stabilizer measurement to highlight the effect of real-time correction. 
Fig.~\ref{fig:fig3}d shows that periodic feedback suppresses fidelity decay and mitigates inter-cycle error propagation. The results closely match those obtained with a single final-round correction using all syndrome data, indicating that repeated mid-circuit feedback introduces no additional degradation while enabling flexible, low-latency real-time control. 
This demonstration establishes that the FPGA-integrated NN decoder and control electronics can dynamically suppress errors during repeated QEC cycles—a prerequisite for more complex logical circuits.

\section{Non-Clifford logical circuit}

\begin{figure}[htbp]
  \centering
  \includegraphics[width=0.5\textwidth]{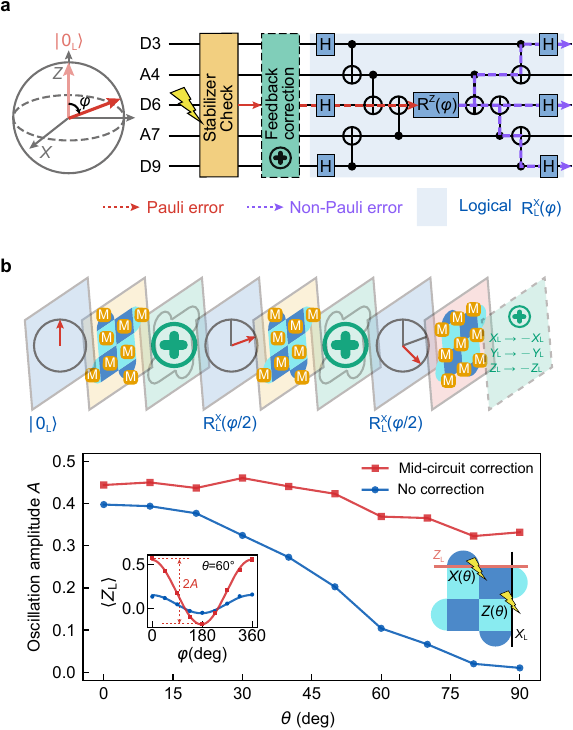}
  \caption{\textbf{Real-time QEC in a non-Clifford logical circuit.}
    \textbf{a}, Circuit (blue background) implementing a logical rotation about the \(X_{\rm L}\) axis. Without feedback correction (green dashed), stochastic Pauli errors (red)—introduced via injected rotation gates (lightning) and subsequent stabilizer measurements (yellow)—can evolve into non-Pauli errors (purple dashed) after the \({\rm R}^Z(\varphi)\) rotation and propagate to multiple data qubits through subsequent entangling gates.  
    \textbf{b}, Experimental demonstration of a non-Clifford logical rotation with mid-circuit QEC (top: experimental circuit). Effective stochastic Pauli \(X\) and \(Z\) errors are generated by applying \(X(\theta)\) and \(Z(\theta)\) rotations on qubits D2 and D6 prior to stabilizer measurements, where \(\theta\) controls the error probability. The net logical rotation \({\rm R}^X_{\rm L}(\varphi)\) is decomposed into two sequential \({\rm R}^X_{\rm L}(\varphi/2)\) gates, with mid-circuit feedback applied before each gate. The logical qubit is initialized in \(|0_{\rm L}\rangle\). 
    The inset shows the expectation value \(\langle Z_{\rm L} \rangle\) oscillating as a function of \(\varphi\), with the oscillation amplitude \(A\) quantifying logical coherence. Logical coherence as a function of \(\theta\) is shown for experiments with (red) and without (blue) mid-circuit feedback correction.
    }
  \label{fig:fig4}
\end{figure}

Pauli-frame updating allows many logical circuits to avoid applying physical correction gates by tracking Pauli byproducts in software. However, this approach fails for non-Clifford operations. For example, under a \(T\) gate, \(T X T^\dagger = (X+Y)/\sqrt{2}\), which is no longer a Pauli operator and cannot be represented within a standard Pauli frame. Consequently, logical circuits involving non-Clifford operations—such as magic-state injection~\cite{bravyi2005universal}, code switching~\cite{daguerre2025code,pogorelov2025experimental}, or related constructions—require mid-circuit feedback to properly account for propagated Pauli byproducts.

We implement a non-Clifford logical rotation about the \(X_{\rm L}\) axis using a simplified, non-fault-tolerant protocol~\cite{marques2022logicalqubit,sun2025logical}. 
The circuit (Fig.~\ref{fig:fig4}a) entangles the logical operator \(X_{\rm L}\) onto physical qubit D6 via CNOT gates, applies a single-qubit \(Z\)-axis rotation \({\rm R}^Z(\varphi)\) on D6, and then disentangles D6. For \(\varphi = 45^\circ\), \({\rm R}^Z(\varphi)\) corresponds to a \(T\) gate. Without mid-circuit feedback correction, a Pauli error occurring before the \({\rm R}^Z(\varphi)\) rotation generally evolves into a non-Pauli error and propagates to other data qubits through the subsequent entangling gates.

To demonstrate real-time QEC with the logical rotation, we decompose the net logical rotation \({\rm R}^X_{\rm L}(\varphi)\) into two sequential \({\rm R}^X_{\rm L}(\varphi/2)\) gates, applying mid-circuit feedback before each gate to correct detected errors.
To illustrate the effectiveness of mid-circuit feedback, Pauli \(X\) and \(Z\) errors are injected before the logical rotations by applying \(X(\theta)\) and \(Z(\theta)\) rotations on qubits D2 and D6 prior to stabilizer measurement. 
Because the direction of the logical rotation depends on the ancilla states, we post-select trajectories in which the \(A4A7\) outcomes retain the same sign across two consecutive rounds of stabilizer measurements, rather than using active or passive qubit reset~\cite{sm2026}. The post-selection serves only to stabilize the logical-rotation demonstration and does not constitute a fundamental requirement for scalable implementations with qubit reset.
By scanning the logical rotation angle \(\varphi\), we observe sinusoidal oscillations of \(\langle Z_{\rm L} \rangle\) (inset of Fig.~\ref{fig:fig4}b). The oscillation amplitude \(A\) serves as a metric of logical coherence: errors in the logical rotation reduce \(A\). We note a small offset in both corrected and uncorrected data, attributed to stronger decoherence when \(\varphi/2 \approx 90^\circ\), i.e., when the logical state after the first rotation lies near the equatorial plane of the Bloch sphere~\cite{sm2026}. 
Across different error probabilities, the mid-circuit feedback preserves significantly larger oscillation amplitudes compared with the uncorrected data, demonstrating improved logical coherence.
These results highlight the necessity and effectiveness of real-time error correction in non-Clifford logical circuits.

\section{conclusion and discussion}

We demonstrate real-time QEC on a distance-3 surface-code logical qubit using an FPGA-based NN decoder tightly integrated with the control hardware.
We use a CFM to manage real-time decoding and distribute feedback instructions to the control electronics, resulting in a deterministic closed-loop latency of 550~ns, well below the 1.25~$\mu$s QEC cycle.
The real-time NN decoder achieves accuracy and robustness comparable to offline MWPM decoder, and decoder-informed feedback correction yields logical error rates nearly identical to offline post-processing.
We further validate real-time error correction in circuits containing non-Clifford gates, where conventional Pauli-frame tracking alone can no longer efficiently account for propagated Pauli byproducts.

These results establish a deterministic hardware-integrated feedback architecture for next-generation QEC experiments, including lattice surgery, magic-state injection, dynamically reconfigured code patches, and future universal fault-tolerant logical operations.
Further scaling to larger code distances and multi-logical-qubit systems will require two hardware improvements: 
(1) Larger code distances require collecting more qubit measurement outcomes at the CFM, imposing stringent demands on low-latency many-to-one communication. Highly integrated DAQ platforms such as RFSoC~\cite{stefanazzi2022qick} can alleviate this bottleneck by enabling a single DAQ to read out more qubits, thereby reducing the total number of DAQ nodes interfacing with the CFM, lowering I/O overhead.
(2) More complex decoders increase FPGA resource consumption. For NN-based decoders, DSP multipliers typically constitute the primary bottleneck owing to the intensive matrix multiplications involved. This constraint tightens as the code distance grows. 
Assuming that the LSTM hidden-layer dimension scales linearly with code distance, current state-of-the-art FPGAs can support NN decoding up to approximately distance-13 surface codes~\cite{sm2026}.
Scaling beyond this regime will likely require higher-end FPGAs or ASICs, multi-FPGA partitioning, DSP time-multiplexing, or more parameter-efficient decoder architectures to preserve the low-latency performance required for real-time decoding.

\bibliography{main}

\section*{Methods}

\subsection{FPGA-based NN decoder design}

LSTM networks \cite{hochreiter1997Long,lecun2015Deep} are a type of recurrent neural network (RNN) that have been shown to provide accurate decoding performance in prior software-based studies~\cite{varbanov2025neural,baireuther2019neural,xin2025Improveda}. Compared with architectures such as Transformers, LSTMs have a more compact structure, making them well suited for implementation on resource-constrained FPGA platforms.

As shown in Fig.~\ref{fig:fig2}a, the error syndromes form an input sequence $\{x_n\}$, which is processed by an LSTM layer with 32 hidden units, yielding hidden and cell states $h_n, c_n \in \mathbb{R}^{32}$. At extraction round $n$, the input $x_n$ and the previous states $h_{n-1}$, $c_{n-1}$ are used to compute the input gate $i_n$, forget gate $f_n$, candidate cell state $\tilde{c}_n$, and output gate $o_n$:
\[
g_n^{(\alpha)} = \mathrm{act}_\alpha\!\left(W_{x}^{\alpha} x_n + W_{h}^{\alpha} h_{n-1} + b^{\alpha}\right), \quad \alpha \in \{i,f,c,o\},
\]
where $g_n^{(\alpha)}$ corresponds to $i_n, f_n, \tilde{c}_n$, and $o_n$, respectively. Here, $W_{x}^{\alpha}$ and $W_{h}^{\alpha}$ denote the input and recurrent weight matrices, and $b^{\alpha}$ are the corresponding bias vectors. The activation function $\mathrm{act}_\alpha$ is a sigmoid-type function for $\alpha \in \{i,f,o\}$ and a ReLU-type function for $\alpha = c$~\cite{nair2010rectified,lecun2015Deep}. The cell state and hidden state are updated as
\[
\begin{aligned}
c_n &= f_n \odot c_{n-1} + i_n \odot \tilde{c}_n,\\
h_n &= o_n \odot \mathrm{ReLU}(c_n),
\end{aligned}
\]
where $\odot$ denotes element-wise multiplication. The states $c_n$ and $h_n$ are propagated to the next iteration, enabling temporal memory.

The hidden state $h_n$ is further processed by a dense layer with a single output neuron, yielding $y_n = \sigma(W_dh_n+b_d)$. A logical flip is declared when $y_n > 0.5$. In the CFM, an $X$-type decoder and a $Z$-type decoder are instantiated to process $X$ and $Z$ logical errors, respectively.

To enable efficient FPGA implementation, the network is quantized using 6-bit signed integer representations.
Nonlinearities are approximated with hardware-friendly functions: the standard tanh activation is replaced by a clipped ReLU, and the sigmoid function is approximated by a piecewise linear form. Both are implemented with saturation to constrain outputs to $[0,1]$:
\[
\begin{aligned}
\mathrm{ReLU}(x) &= \mathrm{clip}(x,0,1) = \min(\max(x,0),1) \\
\sigma(x) &= \mathrm{clip}(0.5x + 0.5,0,1) \\ &= \min(\max(0.5x + 0.5,0),1).
\end{aligned}
\]

\subsection{Closed-loop Latency decomposition}

The total closed-loop latency is decomposed into three contributions:

(1) \textbf{DAQ sampling delay:} The interval between the end of the readout drive pulse and the end of the acquisition window. To capture photons released from the readout resonators, the window is typically extended by $\sim$200~ns beyond the drive pulse. Propagation through several metres of coaxial cables adds a further few tens of nanoseconds. This component is determined by user-defined acquisition settings.

(2) \textbf{Decoding latency on the CFM:} This includes the intrinsic latency of the NN decoder, syndrome extraction (from qubit-state classification to syndrome computation), and the PFU. The NN latency is precisely characterized via digital simulation of the FPGA logic in Vivado~\cite{sm2026}.

(3) \textbf{Control-electronics delays:} These include DAC/ADC conversion, IQ demodulation and qubit-state classification, digital communication, waveform generation in the AWG FPGA, and trigger propagation within the backplane.

A detailed breakdown of each contribution is provided in Table~\ref{tab:latency_breakdown}.

\begin{table*}
  \centering
   \begin{tabular}{l l r}
    \hline\hline
    Category & Contribution & Latency (ns) \\
    \hline
    Readout sampling
      & DAQ sampling time after readout pulses & 222 \\
    \hline
    NN decoder
      & Syndrome calculation & 20 \\
      & Neural network core latency & 124 \\
      & Pauli frame update & 4 \\
      & Subtotal (decoder) & 148 \\
    \hline
    Control electronics (excluding decoder)
      & ADC chip latency~\cite{texasinstruments_adc08d1020_nodate} & 12 \\
      & IQ demodulation in DAQ FPGA & 32 \\
      & Qubit-state classification logic & 4 \\
      & Digital communication & 36 \\
      & Backplane feedback logic latency & 8 \\
      & Trigger propagation from backplane to AWG & 16 \\
      & Waveform-generation logic in AWG FPGA & 32 \\
      & DAC chip latency\cite{analogdevices_ad9739} & 40 \\
      & Subtotal (excluding decoder) & 180 \\
    \hline
    Total closed-loop latency
      &  & 550 \\
    \hline\hline
  \end{tabular}
  \caption{Breakdown of the closed-loop feedback latency at 250~MHz clock rate.}
    \label{tab:latency_breakdown}
\end{table*}

\begin{acknowledgments}
We thank Fei Yan for critical reading of the manuscript.
This work is supported by the National Natural Science Foundation of China (Grants No.~123b2071, No.~12374474, No.~12404582), the Innovation Program for Quantum Science and Technology (Grants No.~2021ZD0301703), the Guangdong Basic and Applied Basic Research Foundation (Grants No.~2024A1515011714). 

\end{acknowledgments}

\section*{Author Contributions}
X.Y., X.S. and Z.W. calibrated the quantum processor. 
X.S. developed the NN decoder and the FPGA for control electronics, collected and analyzed the data. 
J.Z. developed the hardware of control electronics. 
X.S. and J.C. conceived the experiments. 
D.Y. supervised the project. 
All authors contributed to the experimental setup, discussions of the results, and writing of the manuscript.

\section*{Data availability}
The data that support this study are available on Zenodo at https://doi.org/10.5281/zenodo.19908642~\cite{sun_2026_19908642}

\end{document}


\title{Supplementary information for ``Real-time Surface-code Error Correction Using an FPGA-based Neural-Network Decoder''}

\maketitle

\setcounter{equation}{0}
\setcounter{figure}{0}
\setcounter{table}{0}
\setcounter{page}{1}

\tableofcontents
\newpage

\section{The Superconducting Quantum Processor}

\begin{figure*}[htbp]
  \centering
  \includegraphics[width=\textwidth]{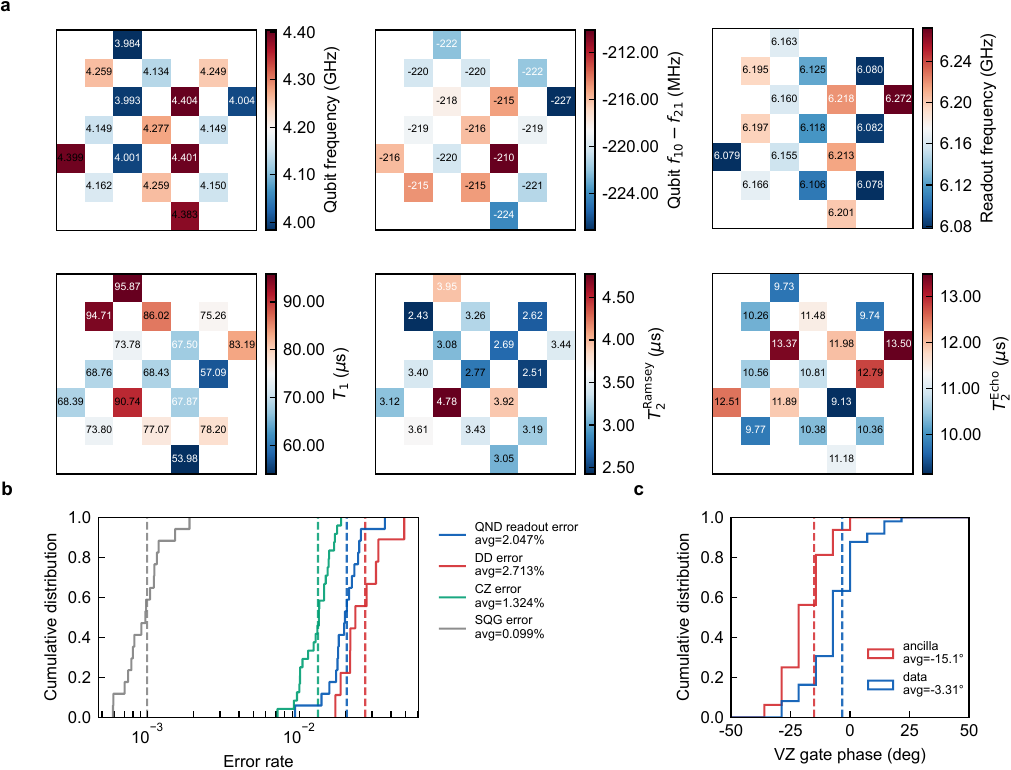}
  \caption{\textbf{Device parameters and gate performance of the 17-qubit subset. }
    \textbf{a}, Heatmaps of various qubit parameters for the distance-3 surface code, including qubit idle frequencies, anharmonicities, frequencies of readout resonators, and coherence times at the operating frequency.
  \textbf{b}, Cumulative distributions of the error rates for parallel single-qubit gates (SQG), CZ gates, QND readout and dynamical decoupling (DD) on the data qubits. 
  \textbf{c}, Rotation angles of the inserted virtual-$Z$ (VZ) gates in the circuit, optimized to reduce the average detection rate per round.
  }
  \label{fig:phy_paramter}
\end{figure*}

The experiments in this work are performed on a 17-qubit subset of a 66-qubit superconducting quantum processor with 110 couplers, which shares the same design as the processor described in Ref.~\cite{sun2025logical}.
The device parameters relevant for implementing a distance-$3$ surface code are summarized in Fig.~\ref{fig:phy_paramter}. Fig.~\ref{fig:phy_paramter}a shows the idle frequencies, anharmonicities, readout resonator frequencies, and coherence times of the 17 physical qubits. Fig.~\ref{fig:phy_paramter}b presents the cumulative distributions of error rates for parallel QND readout, dynamical decoupling (DD), CZ gates, and single-qubit gates required for surface-code operations. The median error rates are $2.05\%$, $2.71\%$, $1.32\%$, and $0.10\%$, respectively. Single-qubit gate errors are characterized via randomized benchmarking (RB), while CZ gate errors are obtained through cross-entropy benchmarking (XEB). The error characterization protocols for DD and QND readout follow Ref.~\cite{sun2025logical}.

To further suppress the average detection rate per round, we compensate for phase errors by inserting virtual-$Z$ (VZ) gates into the circuit, following Refs.~\cite{googlequantumai2021exponential,kelly2015state}. Specifically, a VZ gate is applied before each $H$ gate, and the VZ parameters are optimized to minimize the stabilizer detection rates affected by these phase corrections. Different sets of VZ parameters are used for the first round, intermediate rounds, and the final logical measurement round. The cumulative distributions of the optimized VZ parameters are shown in Fig.~\ref{fig:phy_paramter}c.

\section{Design and Training of Neural Network Decoder}

\subsection{Syndrome Preprocessing}
As shown in Fig.~1b of the main text, the DAQs measure the states of both ancilla and data qubits and forward the outcomes to the central feedback module (CFM). After aggregation, the FPGA performs preprocessing to compute the error syndromes for the current round, which are then used by the decoder.

In the $n$-th round of stabilizer measurement, $a_n^i \in \{0,1\}$ denote the measurement outcome of ancilla qubit $A_i$ and the corresponding stabilizer value $s_n^i$ and error syndrome $x_n^i$ are computed as
\begin{align}
s_n^i &= a_n^i \ \mathrm{XOR}\ a_{n-1}^i, \\
x_n^i &= s_n^i \ \mathrm{XOR}\ s_{n-1}^i.
\end{align}
The XOR between consecutive ancilla outcomes effectively accounts for the reset operation after measurement, which is absent in our circuit.
For the first round ($n = 1$), the system is initialized with $a_0^i = 0$. If the logical state is prepared in the $Z$ basis, the corresponding $Z$-type stabilizers are initialized as $s_0^i = 0~(i=2,4,5,7)$, while the complementary $X$-type stabilizers are set to $s_0^i = s_1^i~(i=1,3,6,8)$; the assignments are reversed if the logical state is prepared in the X basis.

During the final logical measurement, the readout outcomes of data qubits, denoted as $d^i(i=1,2,...,9)$, also generate stabilizers $s_{\mathrm{m}}^i$ and corresponding error syndromes $x_{\mathrm{m}}^i$. For example, in a logical $Z$ measurement,
\begin{align}
s_{\mathrm{m}}^{2} &= d^1 \ \mathrm{XOR}\ d^2 \ \mathrm{XOR}\ d^4 \ \mathrm{XOR}\ d^5, \\
x_{\mathrm{m}}^{2} &= s_n^2 \ \mathrm{XOR}\ s_{m}^2,
\end{align}
where $x_{\mathrm{n}}^{2}$ is the stabilizer value from the last round of ancilla measurements.

As discussed in the main text, active feedback corrections can introduce artificial syndromes in subsequent rounds. 
For instance, applying an $X$ gate on $D_1$ corrects a logical flip error (recall $Z_L = Z_1 Z_2 Z_3$); however, this gate anticommutes with the stabilizer $Z_1 Z_3 Z_4 Z_5$ associated with $A_2$, causing a flip in the syndrome $x_{n+1}^2$. To prevent these correction-induced signals from being misinterpreted as physical errors, the preprocessing module conditionally applies an additional $\mathrm{XOR}\ 1$ to $x_{n+1}^2$, canceling the artificial syndrome.

All variables are stored in FPGA registers, enabling fully pipelined operation. The total latency of the preprocessing stage for syndrome calculation is fixed at $20\,\mathrm{ns}$ (5 FPGA clock cycles).

\subsection{Neural Network Architecture and Hyperparameters}

\begin{figure*}[htb]
  \centering
  \includegraphics[width=\textwidth]{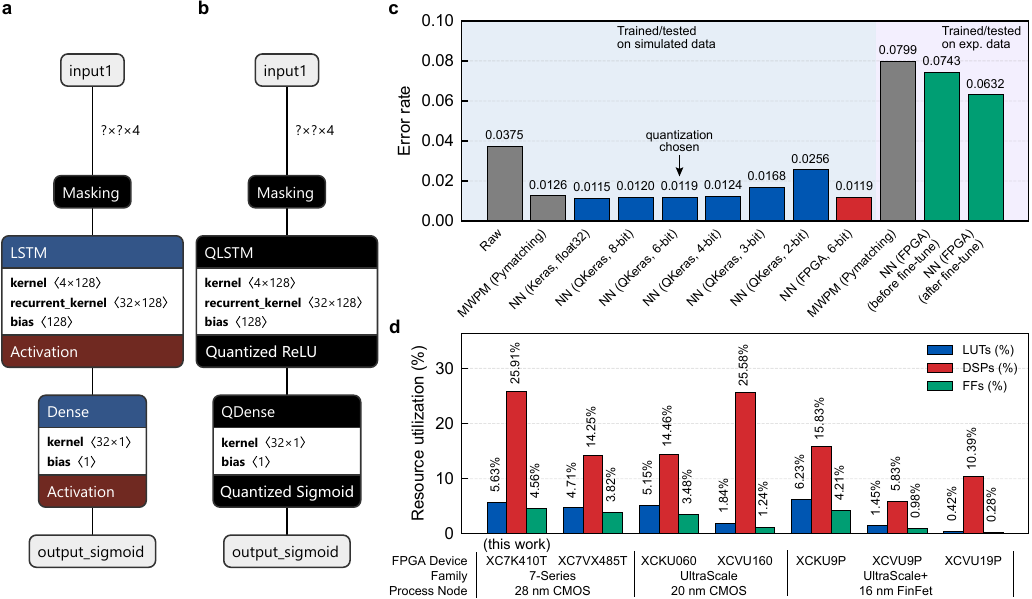}
  \caption{\textbf{Design and training of neural network (NN) decoder.} 
\textbf{a}, Schematic of the non-quantized NN. 
\textbf{b}, Schematic of the quantized NN. 
\textbf{c}, Error rates of different models during pre-training and fine-tuning. Blue-shaded regions correspond to simulated data generated by STIM, and pink-shaded regions correspond to experimental data. The performance of the minimum-weight perfect matching (MWPM) decoder is shown for reference.
\textbf{d}, Resource utilization of the $X/Z$ decoder synthesized on various AMD FPGAs. Red, blue, and green bars denote DSP slices, look-up tables (LUTs), and flip-flops (FFs), respectively.}
\label{fig:fpga_decoder}
\end{figure*}

The neural network (NN) consists of a single LSTM layer followed by a fully connected dense output layer (Fig.~\ref{fig:fpga_decoder}a), designed using the Keras API~\cite{chollet2015keras} with a TensorFlow backend~\cite{abadi2016TensorFlow}. A quantized version, with identical architecture but quantized parameters, is shown in Fig.~\ref{fig:fpga_decoder}b, implemented with QKeras~\cite{coelho2020ultra}. The LSTM network contains 4,736 weight parameters, including kernels and biases (dimensions shown in Fig.~\ref{fig:fpga_decoder}a and b).

For simplicity during training, each input syndrome sample is formatted as \(1 \times 20 \times 4\), where 4 represents the \(Z\) (or \(X\)) stabilizers of the distance-3 surface code and 20 corresponds to QEC rounds.
If fewer than 20 rounds are present, remaining entries are padded with \(-1\) and an additional masking layer is added to prevent these \(-1\) values from propagating to the subsequent LSTM layers. 


\subsection{Training Strategy and Quantization}

Training of the NN decoder proceeds in two stages: pre-training on simulated data generated by STIM~\cite{gidney2021Stim}, followed by fine-tuning on experimental data. In pre-training, STIM produces 2{,}000{,}000 memory-circuit samples with varying numbers of rounds, of which 80\% are used for training and 20\% for testing. Training hyperparameters are summarized in Table~\ref{tab:training_settings}.
During fine-tuning, 100{,}000 experimental samples are used to train and validate the NN, employing the same training settings. Noise characteristics differ between the simulated and experimental datasets.

Initially, a non-quantized (float32) model is trained to provide a reference for the optimal error rate (Fig.~\ref{fig:fpga_decoder}c). Models with different weight bit-widths are then trained; “6-bit” indicates that model weights are stored in a 6-bit format. Using too few bits reduces accuracy, while excessive bit-width increases FPGA resource consumption. Based on these results, we adopt 6-bit quantization for FPGA deployment, achieving error rates comparable to the QKeras model.

The trained weights can be hot-updated into the FPGA’s on-chip RAM, and fine-tuning on experimental data further improves performance. All training is performed on a single NVIDIA RTX 4080 GPU with 16~GB of memory.

\begin{table}[htbp]
\begin{tabular}{p{0.32\linewidth} p{0.62\linewidth}}
\hline
\textbf{Item} & \textbf{Setting / Description} \\
\hline
Task type & Binary classification \\
Optimizer & Adam \\
Initial learning rate & $1 \times 10^{-3}$ \\
Loss function & Binary cross-entropy \\
Metric & Accuracy \\
Batch size & 256 \\
Maximum epochs & 50 \\
EarlyStopping & Monitor: \texttt{val\_loss}; patience: 5\\
\hline
\end{tabular}
\centering
\caption{Training hyperparameters and settings.}
\label{tab:training_settings}
\end{table}

\subsection{FPGA Resource Utilization Analysis}
Fig.~\ref{fig:fpga_decoder}d shows the resource utilization of the $X/Z$ decoders on different AMD FPGAs~\cite{xilinx_7_2020}. To accelerate the matrix operations in the NN, a large number of DSP multipliers are used, making the DSP slices the primary limiting resource. In this work, we implement the decoder on a mid-range AMD FPGA (XC7K410T), where each decoder occupies 25.91\% of the available DSPs, 5.63\% of the LUTs, and 4.56\% of the FFs. The current NN size is therefore well below the FPGA resource limit.

For surface codes with larger code distances, larger NN decoders are required. In such cases, FPGAs with more abundant resources (such as XCVU9P) can be used, and techniques such as reusing matrix-multiplication modules can further reduce hardware resource consumption.

\begin{table}[t]
\begin{tabular}{cccccc}
\toprule
$d$ & $\dim(x)$ & $h$ & $P_{\mathrm{LSTM}}$ & DSP (\% of VU13P) & Latency (ns) \\
\midrule
3  & 4   & 32  & 4{,}736    & 399 (3.2\%)        & 124 \\
5  & 12  & 53  & 13{,}992   & 1{,}094 (8.9\%)    & 205 \\
7  & 24  & 75  & 29{,}700   & 2{,}191 (17.8\%)   & 291 \\
9  & 40  & 96  & 52{,}224   & 3{,}591 (29.2\%)   & 372 \\
11 & 60  & 117 & 83{,}304   & 5{,}337 (43.4\%)   & 453 \\
13 & 84  & 139 & 123{,}212  & 7{,}533 (61.3\%)   & 538 \\
15 & 112 & 160 & 172{,}160  & 9{,}975 (81.2\%)   & 620 \\
17 & 144 & 181 & 230{,}364  & 12{,}757 (103.8\%) & 701 \\
\bottomrule
\end{tabular}
\centering
\caption{
First-order estimates of per-decoder resources and inference latency for a hypothetical deployment on an AMD VU13P, extrapolated from the $d=3$ implementation. $\dim(x)$ is the input dimension per round, $h$ is the LSTM hidden-layer size, $P_{\mathrm{LSTM}}$ is the total number of LSTM parameters, DSP usage is expressed both in absolute slices and as a percentage of the VU13P resources, and Latency is the estimated inference time per round.
}
\label{tab:scaling}
\end{table}

\subsection{Resource Estimation for Decoders at Larger Code Distances}
\label{sec:scaling}

To assess the scalability of the proposed architecture, we provide a simple scaling model anchored to our $d=3$ implementation, projecting the FPGA resource consumption and inference latency of the LSTM decoder as the code distance $d$ increases.

For a rotated surface code, each stabilizer type (X or Z) is associated with $(d^2-1)/2$ ancilla qubits, so the per-round input dimension scales as $\dim(x) = (d^2-1)/2 \propto d^2$. Prior LSTM-based decoders scaled the hidden-layer size approximately linearly with code distance ($h = 64, 96, 128$ for $d = 3,5,7$)~\cite{varbanov2025neural}, achieving logical performance superior to that of the MWPM decoder.
We therefore adopt $h \propto d$ as a working assumption. A single-layer LSTM with input dimension $\dim(x)$ and hidden size $h$ contains $P_{\mathrm{LSTM}} = 4(\dim(x),h + h^2 + h) \propto d^3$ weight parameters.

In FPGA design, the parameter count $P_{\mathrm{LSTM}} \propto d^3$ sets the total resource-time product, $\mathrm{DSP} \times \mathrm{Latency} \propto d^3$. To preserve the parallelization strategy of our $d=3$ implementation, we assume $\mathrm{DSP} \propto h^2 \propto d^2$, and the remaining factor then yields $\mathrm{Latency} \propto d$. Anchored to the measured 399 DSPs and 124\,ns per decoder at $d=3$, these scalings give the projections summarized in Table~\ref{tab:scaling}.

Our current prototype is deployed on a mid-range AMD Kintex-7 410T (1{,}540 DSP slices). Extrapolating to a high-end VU13P with 12{,}288 DSP slices, a single FPGA can support a decoder up to $d\approx 13$ (61\% DSP usage, 538\,ns latency), with the resource limit reached near $d=15$. Scaling beyond this point would require multi-FPGA partitioning, DSP time-multiplexing, or more parameter-efficient decoder architectures.

We emphasize that these estimates are first-order projections rather than precise predictions. 
The model relies on simplifying assumptions---most notably that parallelism can be scaled in step with the hidden size, and that the number of LSTM layers remains fixed across distances---which may not hold in practice. Factors such as BRAM capacity, routing congestion, timing closure, and achievable logical error rate at each distance must be evaluated case-by-case in concrete hardware implementations.

\section{Performance of Real-time NN decoder}

\begin{figure*}[htbp]
  \centering
  \includegraphics[width=\textwidth]{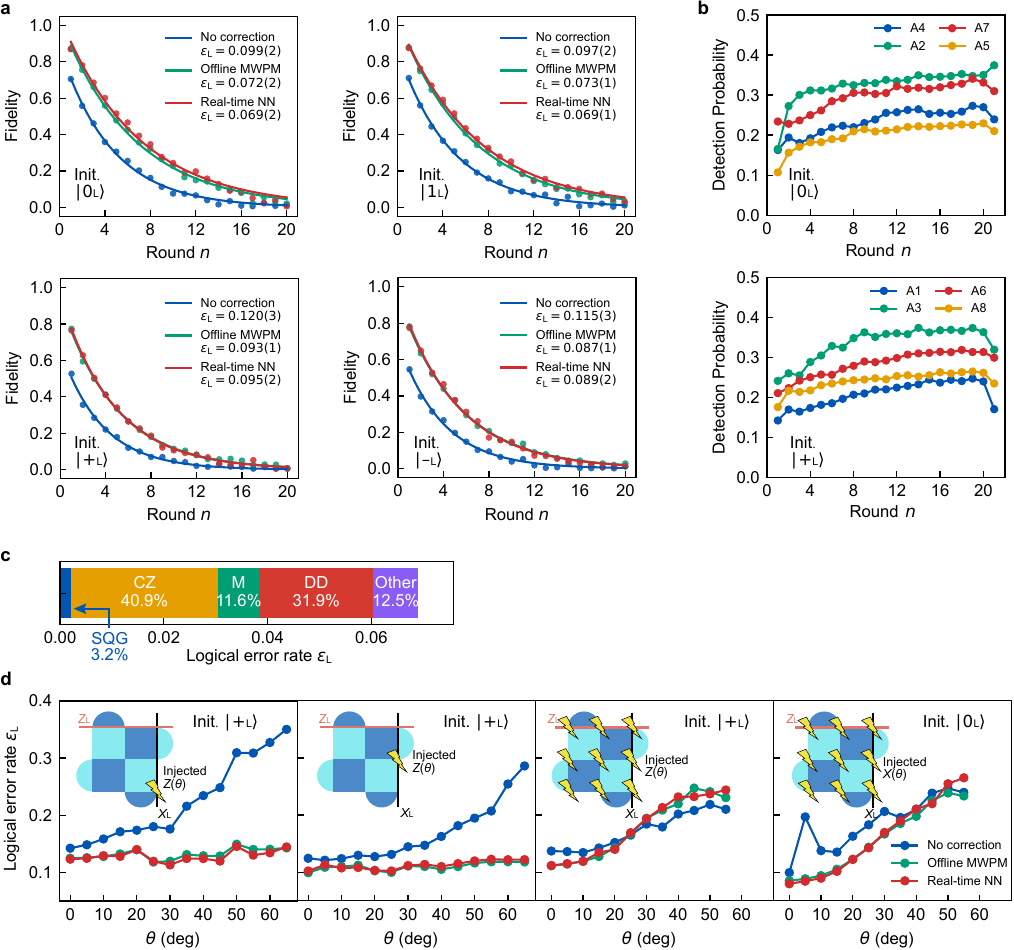}
  \caption{\textbf{Performance of the NN Decoder.}
\textbf{a}, Logical memory results for input states \(\ket{0_L}\), \(\ket{1_L}\), \(\ket{+_L}\), and \(\ket{-_L}\).
Errors are corrected using the offline MWPM decoder (green) and the real-time NN decoder (red), implemented via software- and FPGA-based FPU, respectively.
\textbf{b}, Error detection probabilities recorded by the CFM for initial states \(\ket{0_L}\) (top) and \(\ket{+_L}\) (bottom).
\textbf{c}, Error budget for the logical $\lvert 0_L \rangle$ state. The total logical error is $6.9(2)\%$, with contributions from CZ gates, single-qubit gates (SQG), dynamical decoupling (DD), ancilla measurement (M) and other sources.
\textbf{d}, Decoder robustness under various error injections.
Left two panels show logical error rates for \(\ket{+_L}\), with \(Z\) error injected on D6 and D9 separately before each QEC round; the angle \(\theta\) sets the physical error probability.
Right two panels show logical error rates with \(Z\) and \(X\) errors injected on all data qubits, implemented via virtual-Z (VZ) gates before (for \(Z\)) or after (for \(X\)) the first \(H\) gate in each round.
}
\label{fig:exp01_memory}
\end{figure*}

Here we present additional data complementing Fig.~2 in the main text, summarized in Fig.~\ref{fig:exp01_memory}.

Fig.~\ref{fig:exp01_memory}a shows logical memory experiments for the four logical states \(\ket{0_L}\), \(\ket{1_L}\), \(\ket{+_L}\), and \(\ket{-_L}\). 
In these experiments, we employ Pauli frame update (PFU) instead of real-time feedback to isolate the performance of the decoder itself. For all four logical states, logical error rates per round obtained with the real-time NN decoder closely match those from the offline MWPM decoder.

Fig.~\ref{fig:exp01_memory}b presents the error detection probability in the \(\ket{0_L}\) and \(\ket{+_L}\) memory circuits, with the detection events recorded by the CFM. The detection probability increases with the number of rounds, reflecting the gradual accumulation of leakage~\cite{mcewen2021removing} and indicating that dedicated leakage-reduction operations~\cite{huber2025parametric,lacroix2025fast,marques2023allmicrowave,miao2023overcoming,yang2024CouplerAssistedb,he2025experimental} will be required in future implementations.

Fig.~\ref{fig:exp01_memory}c shows the error budget for $\lvert 0_L \rangle$. The total logical error rate is $6.9(2)\%$, with dominant contributions from CZ gates and dynamical decoupling (DD).

Fig.~\ref{fig:exp01_memory}d illustrates the decoder's robustness under various noise injections.
The two left panels show logical error rates when \(Z\) noise is injected on D6 and D9, demonstrating that the NN decoder reliably suppresses logical errors. 
The two right panels correspond to \(Z\) and \(X\) noise applied to all data qubits.
When the noise-injection angle becomes sufficiently large, the corrected logical error rate can exceed the raw error rate without error correction, reflecting the fact that a distance-3 surface code only guarantees protection against single-qubit errors. In all scenarios, the real-time NN decoder closely matches the performance of the offline MWPM decoder, demonstrating robustness even under conditions outside the original NN training distribution.

\begin{figure*}[htbp]
  \centering
  \includegraphics[width=\textwidth]{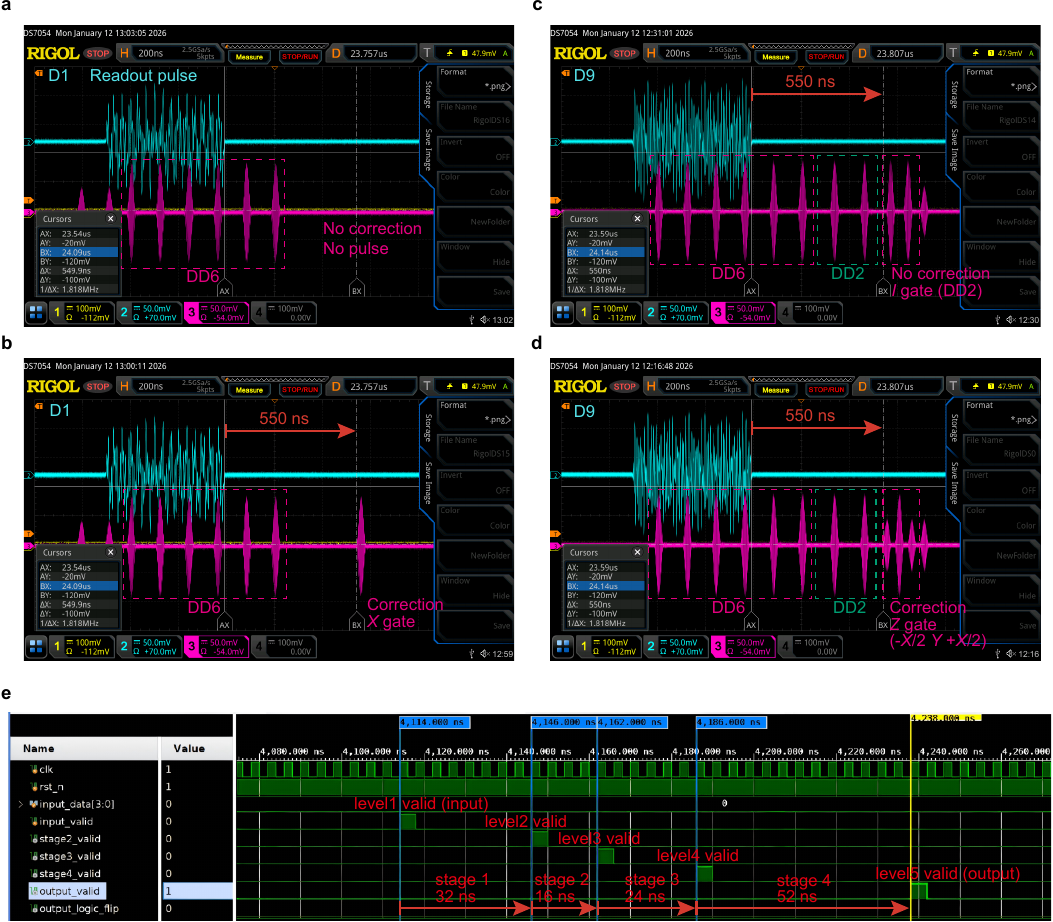}
  \caption{\textbf{Calibration of Feedback Latency.}
\textbf{a}, 
Oscilloscope traces of the simultaneous readout signals for ancilla qubits (sky-blue) and drive signals for data qubit D1 (pink).
The waveforms correspond to a logical memory experiment for \(\lvert 0_L \rangle\) without mid-circuit logical \(Z_L\) correction.
During ancilla measurement (including the post-delay period for resonator photon depletion), a dynamical decoupling sequence with 6 $X$ pulses (DD6) is applied to D1 to mitigate dephasing errors.
\textbf{b}, Same as (a), but with an additional \(X\) gate applied on D1 to implement a logical \(Z_L\) flip.
\textbf{c}, Waveforms for a logical \(\lvert +_L \rangle\) memory experiment without applying a logical \(X_L\) correction. 
A dynamical decoupling sequence with two $X$ pulses (DD2), also serving as an idle ($I$) gate, suppresses dephasing errors during the feedback waiting period. 
\textbf{d}, Same as (c), but with a logical \(X_L\) correction applied. 
A physical \(Z\) gate, implemented via the sequence (\(-X/2\), \(Y\), \(+X/2\)) on D9, flips \(X_L\).
\textbf{e}, Digital simulation of the NN decoder core, with the latency of each stage is indicated. Stage definitions are provided in Fig.~2a of the main text.}
  \label{fig:vivado_sim}
\end{figure*}

\section{Closed-Loop Feedback Latency Calibration}

The closed-loop latency is defined as the time delay between the end of the readout drive pulse and the beginning of the feedback pulse. 
To measure this latency, the output port used for the ancilla-qubit readout drive and the feedback-control AWG output are connected to an oscilloscope through coaxial cables of equal length. 
Fig.~\ref{fig:vivado_sim}a–d show examples with and without logical \(Z_L\) and \(X_L\) corrections. 
These measurements indicate a stable closed-loop latency of approximately 550~ns, which can be decomposed into three components:
(1) \textbf{DAQ sampling delay:} The time between the end of the readout drive pulse and the end of the DAQ sampling window.
To fully capture photons released from the readout resonator, the sampling window is typically set $\sim$200~ns longer than the drive pulse. Additional propagation delays from several meters of coaxial cables introduce a further tens of nanoseconds. This delay is determined by user-defined acquisition settings.
(2) \textbf{Decoding latency on the CFM:} This includes the intrinsic NN latency, the latency of syndrome calculation (from qubit-state classification to syndrome extraction), and the latency associated with the PFU. The NN decoder latency is precisely characterized via digital simulation of the FPGA logic in Vivado (see Fig.~\ref{fig:vivado_sim}e).
(3) \textbf{Control-electronics delays:} These include DAC/ADC conversion, IQ demodulation and qubit-state classification, digital communication, waveform-generation logic in the AWG FPGA, and trigger propagation within the backplane.

\section{Real-time QEC}

\subsection{Logical memory experiment with feedback correction}

\begin{figure*}[htbp]
  \centering
  \includegraphics[width=\textwidth]{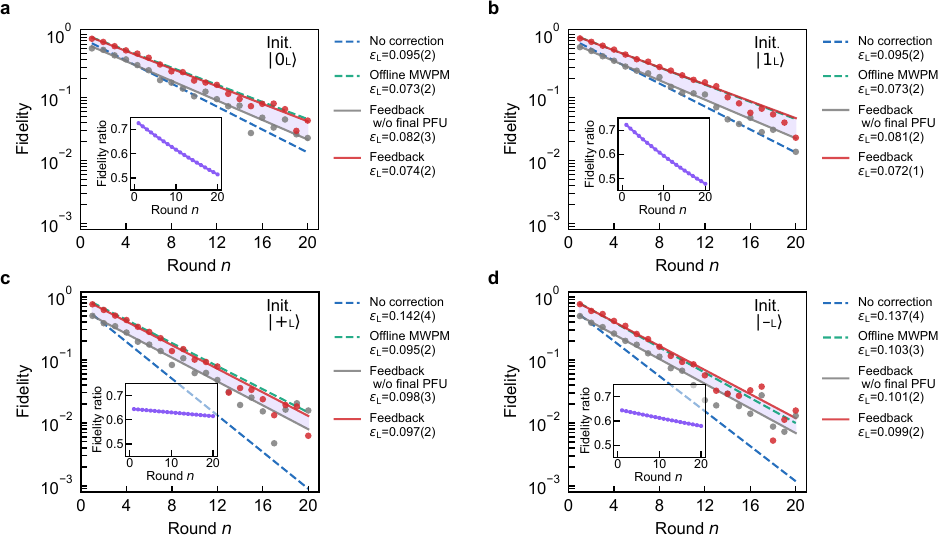}
  \caption{\textbf{Logical memory experiment with feedback correction}. Logical fidelities versus QEC round \(n\) with final-round feedback for initial states \(\lvert 0_L \rangle\), \(\lvert 1_L \rangle\), \(\lvert +_L \rangle\), and \(\lvert -_L \rangle\). 
  Grey dots indicate results with a final-round feedback correction before the final logical measurement, while red dots show the case with additional PFU applied after the logical measurement. Insets display the ratio of fidelities without and with PFU as a function of the number of QEC rounds.}
  \label{fig:fb_once}
\end{figure*}

Fig.~\ref{fig:fb_once} shows the logical memory performance for the states \(\lvert 0_L \rangle\), \(\lvert 1_L \rangle\), \(\lvert +_L \rangle\), and \(\lvert -_L \rangle\).
A feedback correction is inserted between the last QEC round and the final logical measurement (grey markers). 
We can further use the stabilizer outcomes obtained during the logical measurement to correct additional errors (red markers) via the final PFU in the FPGA. 
Including the final PFU leads to a clear improvement in logical-state fidelity, with performance closely matching that of the offline MWPM decoder used as a reference (green curves).

The insets display the ratio of fidelities without and with the final PFU, which decreases as the number of QEC cycles grows. This indicates that the PFU applied after the logical measurement effectively corrects both fixed errors (e.g., readout errors during the logical measurement) and errors accumulated over multiple cycles. The different decay rates of the fidelity ratio along the logical \(Z\) and \(X\) directions likely reflect distinct leakage dynamics of the corresponding data qubits, as well as imbalanced logical lifetimes between \(Z\)- and \(X\)-basis states.

\subsection{Logical memory experiment with multi-round feedback}

\begin{figure*}[htbp]
  \centering
  \includegraphics[width=0.95\textwidth]{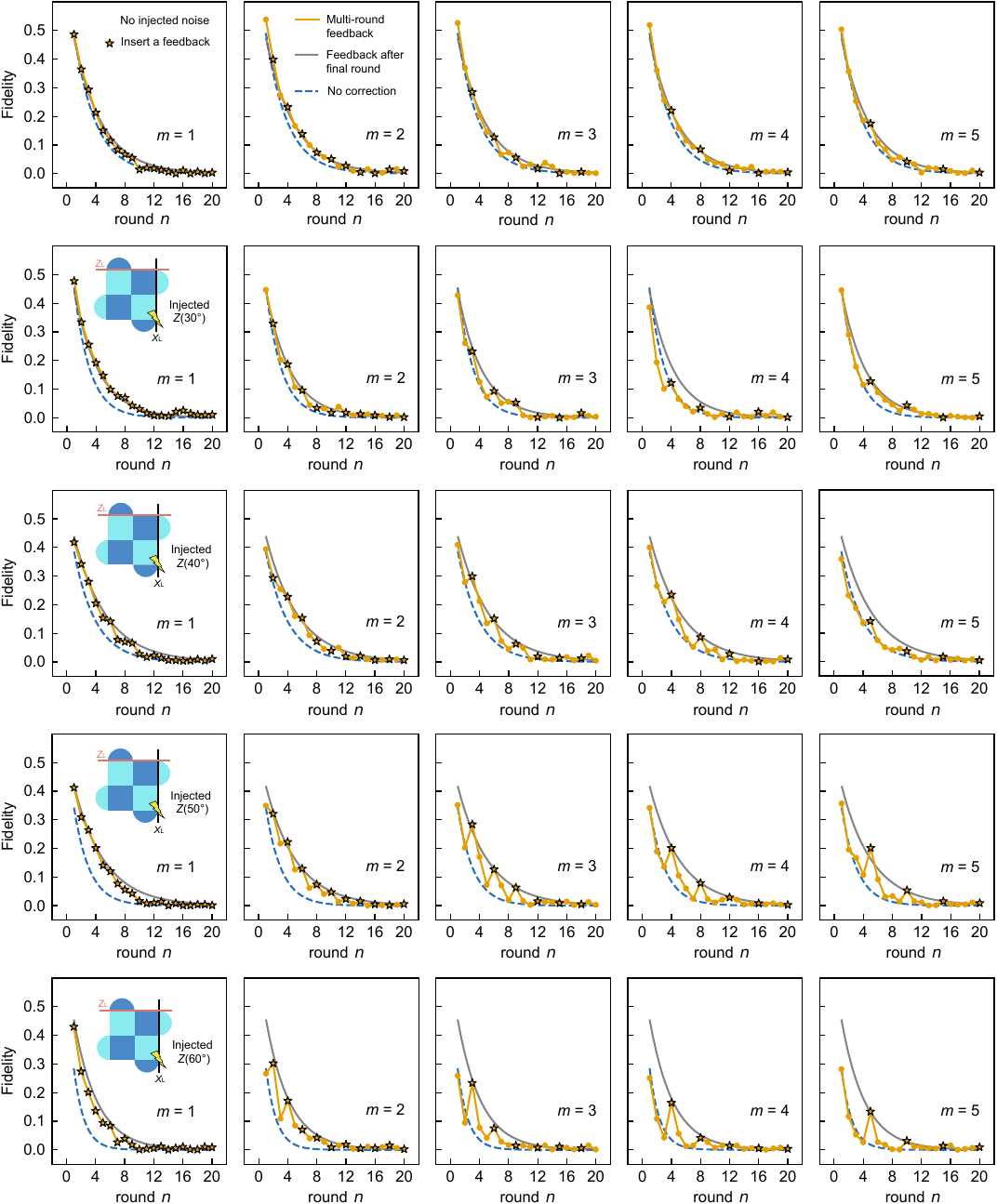}
  \caption{\textbf{Logic memory experiment with multi-round feedback.}
    Performance with feedback correction applied every \(m\) rounds. The logical state is initialized in \(\lvert +_L \rangle\).
    Rows from top to bottom correspond to increasing artificial noise on data qubit D9, with noise amplitude set by physical \(Z\)-rotation angles of 0\(^{\circ}\), 30\(^{\circ}\), 40\(^{\circ}\), 50\(^{\circ}\), and 60\(^{\circ}\). 
    Orange stars indicate rounds where feedback correction is applied. 
    The dashed-blue line shows uncorrected data, and the solid-grey line shows data with final-round feedback correction using all syndrome information. No PFU after the logical measurement is added for all data.
    }
  \label{fig:fb_multir}
\end{figure*}

We investigate logical memory performance using multi-round feedback under varying artificial rotation noise strengths.
The logical state is initialized in \(\lvert +_L \rangle\), and artificial rotation noise is applied to data qubit D9 in each QEC round, with the amplitude set by the physical \(Z\)-rotation angle.
Fig.~\ref{fig:fb_multir} presents results both in the absence of artificial noise and for rotation angles ranging from 30\(^{\circ}\) to 60\(^{\circ}\).

Each orange star marker denotes a feedback correction applied after the corresponding QEC round. The dashed-blue and solid-grey lines serve as reference traces: the dashed-blue line represents uncorrected data, while the solid-grey line corresponds to a final-round feedback correction before the logical measurement. No PFU is applied after the logical measurement in these experiments.

In the absence of artificial noise, the difference between the reference traces is small due to the limited code distance (\(d=3\)), and the benefit of feedback is partially obscured by statistical fluctuations. For nonzero artificial noise, however, each inserted feedback correction restores the logical fidelity toward the uncorrected reference line, demonstrating that mid-circuit feedback effectively preserves logical-state fidelity.

\section{Non-Clifford Logical Circuit}

\begin{figure*}[htbp]
  \centering
  \includegraphics[width=\textwidth]{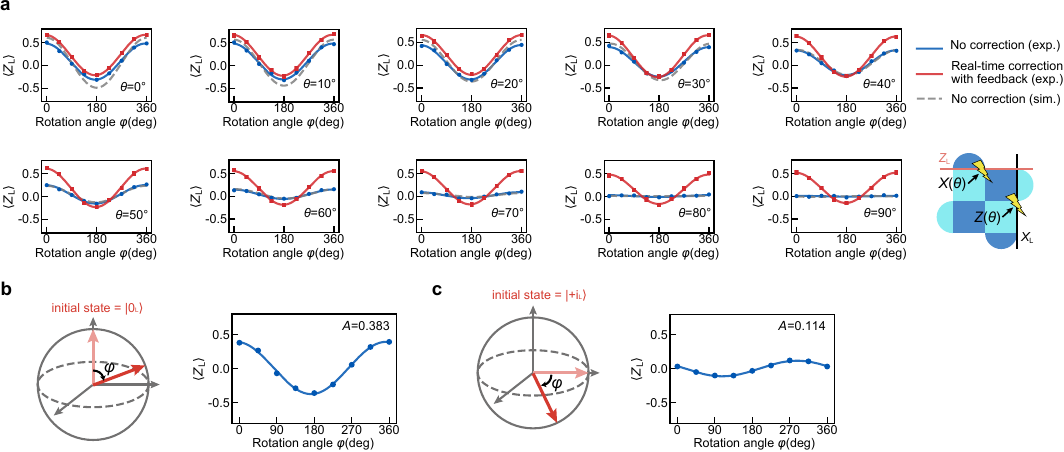}
  \caption{\textbf{Non-Clifford logical rotations with mid-circuit feedback correction.}
\textbf{a}, Logical rotation around the \(X_{\rm L}\) axis for the initial state \(\lvert 0_L \rangle\), obtained by consecutively applying two logical rotation gates with angle \(\varphi/2\). 
Artificial \(X\) and \(Z\) errors are injected on data qubits D2 and D6 before each rotation, with error probabilities set by \(\theta\).
Red and blue traces show data with and without mid-circuit feedback correction, respectively, while grey dashed lines represent noisy simulations without feedback.
\textbf{b,c}, Single logical rotation around \(X_{\rm L}\) without feedback correction, starting from \(\lvert 0_L \rangle\) (\textbf{b}) and \(\lvert +i_L \rangle\) (\textbf{c}). 
One QEC round is performed before the rotation, with no artificial error injection.
}
\label{fig:rotation}
\end{figure*}

Fig.~\ref{fig:rotation}a shows oscillations obtained by consecutively applying two logical rotation gates with angle \(\varphi/2\). Artificial \(X\) and \(Z\) errors are injected on data qubits D2 and D6 before each rotation, with error probabilities set by \(\theta\).
The ten subpanels correspond to the ten data points in Fig.~4b of the main text.
As the error probability increases, the oscillation amplitude \(A\) of the raw data (blue) decreases, whereas data with feedback corrections (red) largely preserve \(A\), demonstrating that real-time feedback correction effectively protects logical coherence in non-Clifford circuits.
Grey dashed lines depict noisy simulations without feedback correction using Cirq~\cite{CirqDevelopers_2025}.

In the non-Clifford circuit of Fig.~4a (main text), data qubits D3, D6, and D9 use ancilla-assisted entanglement via A4 and A7 to load \(X_L\) onto D6, followed by a physical rotation gate. 
The \(A4A7\) outcomes from the previous parity checks determine the direction of subsequent logical rotation. 
Instead of resetting the ancilla qubits to \(|0\rangle\) after each stabilizer measurement~\cite{google2025quantum,he2025experimental}, we post-select data where the \(A4A7\) outcomes are consistent across two parity-check rounds. This ensures the two logical rotations do not cancel each other. Since \(A4A7\) corresponds to the product of two stabilizers, \((Z3Z6)(Z5Z6Z8Z9)=Z3Z5Z8Z9\), this post-selection also partially suppresses errors occurring on data qubits D3, D5, D8, and D9. The same post-selection procedure is applied consistently to all datasets, with and without feedback, and does not affect the conclusions.

All oscillation curves in Fig.~\ref{fig:rotation}a exhibit a noticeable offset: the amplitude \(A\) is larger when \(\varphi\) is close to \(0^\circ\) or \(360^\circ\), and smaller when \(\varphi\) is close to \(180^\circ\). This is attributed to stronger decoherence when \(\varphi/2 \approx 90^\circ\), i.e., when the logical state after the first rotation lies near the the Bloch-sphere equator. To verify, we perform single logical rotations around the \(X_{\rm L}\) axis without feedback correction, initializing the state in \(\lvert 0_L \rangle\) (Fig.~\ref{fig:rotation}b) and \(\lvert +i_L \rangle\) (Fig.~\ref{fig:rotation}c), respectively. A QEC round is inserted before the logical operation with no artificial error injection. The oscillation amplitude for \(\lvert +i_L \rangle\) is smaller than that for \(\lvert 0_L \rangle\), confirming stronger decoherence near the equator. We anticipate that using an XZZX surface code~\cite{bonillaataides2021xzzx} could balance logical lifetimes between the \(Z\)- and \(X\)-basis logical states.

\clearpage
\bibliographystyle{naturemag}
\bibliography{supp}  